# Selection rules in symmetry-broken systems by symmetries in synthetic dimensions

Matan Even Tzur[1], Ofer Neufeld[1], Avner Fleischer[2], and Oren Cohen[1]

[1] Technion – Israel Institute of Technology, 3200003 Haifa, Israel

[2] Raymond and Beverly Sackler Faculty of Exact Science, School of Chemistry and Center for Light-Matter-Interaction, Tel Aviv University, Tel-Aviv 6997801, Israel

**Selection rules are often considered a hallmark of symmetry. When a symmetry is broken, e.g., by an external perturbation, the system exhibits selection-rule deviations which are often analyzed by perturbation theory. Here, we employ symmetry-breaking degrees of freedom as synthetic dimensions, to demonstrate that symmetry-broken systems systematically exhibit a new class of symmetries and selection rules. These selection rules determine the scaling of a system's observables (to all orders in the strength of the symmetry-breaking perturbation) as it transitions from symmetric to symmetry-broken. We specifically analyze periodically driven (Floquet) systems subject to two driving fields, where the first field imposes a spatio-temporal symmetry, and the second field breaks it, imposing a symmetry in synthetic dimensions. We tabulate the resulting synthetic symmetries for (2+1)D Floquet group symmetries and derive the corresponding selection rules for high harmonic generation (HHG) and above-threshold ionization (ATI). Finally, we observe experimentally HHG selection rules imposed by symmetries in synthetic dimensions. The new class of symmetries & selection rules extend the scope of existing**

**symmetry breaking spectroscopy techniques, opening new routes for ultrafast spectroscopy of phonon-polarization, spin orbit coupling, and more.**

Selection rules are a significant consequence of symmetries, appearing throughout science. For example, point-group symmetries forbid electronic transitions in solids and molecules[1], existence of electric and magnetic moments[2], and harmonic generation in perturbative nonlinear optics[3]. In periodically driven (Floquet) systems, symmetries that involve space and time (denoted dynamical symmetries (DSs)) result in selection rules for various phenomena. Examples include symmetry-induced dark states and transperancy[4,5], forbidden harmonics and polarization restrictions in HHG[6–8], and forbidden photoionization channels[9]. Symmetry and selection rules are widely used for controlling and measuring one another. On one hand, a system's symmetry is often "engineered" to achieve a desirable selection rule. A salient example for "selection-rule-engineering" is the generation of circularly polarized XUV emission by discrete rotational symmetries in HHG[10–15]. On the other hand, in symmetry-broken systems the selection rules are replaced with selection rule *deviations* that can be used to extract information about the system and the symmetry breaking perturbation. In nonlinear optics for instance, selection rule deviations are used as a background free gauge of molecular symmetry[16], orientation[17], and chirality[18–20], circulating electric currents[21], Berry curvature[22], topological phase transitions[23], and more[24]. However, selection rule deviations have only been analyzed ad hoc[16–24], mostly via perturbation theory (which is often limited to analyzing selection rule deviations up to $2^{nd}$ order). As a result, there is limited insight about the interplay between selection rule deviations, the broken symmetry, and the symmetry-breaking perturbation.

Here we present a new class of symmetries and selection rules in systems that exhibit broken symmetries in space and time by exploiting the symmetry-breaking degrees of freedom as synthetic dimensions. We term the new class of symmetries "real-synthetic symmetries" because they are composed of a broken symmetry operation that acts in real spatio-temporal dimensions (denoted $\hat{X}$) and a synthetic-dimensions operation (denoted $\breve{\zeta}$). Only the composite operation ($\hat{X} \cdot \breve{\zeta}$), is a symmetry of the physical system (i.e., it does not exhibit $\hat{X}$ and $\breve{\zeta}$ as separate symmetries). The selection rules that result from these synthetic symmetries determine analytically the allowed/forbidden deviations from the selection rules of the fully-symmetric and perturbation free system's expectation values, in a non-perturbative manner. That is, real-synthetic symmetries analytically provide system-independent scaling laws of selection rule deviations in symmetry-broken systems, valid to all orders in the strength of the perturbation, without invoking perturbation theory at any stage. This is in sharp contrast with previous investigations, where deviations from selection rules were only analyzed ad hoc[16–24] using 1st or 2nd order perturbative arguments. Whereas real-synthetic symmetries and their selection rules are general concepts applicable to all symmetry broken systems, we focus here on periodically driven (Floquet) systems subject to two driving fields. The first field imposes a dynamical symmetry $\hat{X}$ (tabulated within Floquet group theory[8]). The second field breaks $\hat{X}$, and instead, imposes a real-synthetic symmetry $\hat{X} \cdot \breve{\zeta}$ where $\breve{\zeta}$ operates on the synthetic space spanned by polarization components of the second field. We employ the operations $\hat{X} \cdot \breve{\zeta}$ to derive selection rules for HHG, ATI, and other phenomena. Finally, we experimentally investigate these selection rules by driving HHG with bi-elliptical pumps with incommensurate frequencies. In our experiment, we observe two types of selection rules – standard HHG selection rules (forbidden harmonics and polarization restrictions) imposed by dynamical (spatio-temporal) symmetries, and selection rules rooted in synthetic dimensions that manifest as scaling laws of the harmonic powers and polarization states as the system transitions from the symmetric state to the strongly perturbed state.

We start by considering a general Floquet system with period $T = 2\pi/\omega$, and a DS denoted by $\hat{X}$. The operation $\hat{X}$ is a (2+1)D spatio-temporal symmetry, jointly imposed by the symmetries of the target material and the first driving laser (or by any other periodic excitation of the system[25,26]). The operations $\hat{X}$ were comprehensively tabulated within the framework of Floquet group theory[8]. Explicitly, the Hamiltonian of the system exhibits a discrete time-translation symmetry $\hat{H}_0(t) = \hat{H}_0(t + T)$ as well as $[\hat{\mathcal{H}}_f, \hat{X}] = 0$ where $\hat{\mathcal{H}}_f \equiv \hat{H}_0 - i\partial_t$ is the Floquet Hamiltonian.

Next, we consider dressing the system with an additional laser with electric field polarization vector $\boldsymbol{p} = (p_x, p_y)$, and frequency is $s\omega = 2\pi s/T$. The Hamiltonian of the dressed system is given by

$$\hat{H} = \hat{H}_0 + \hat{W} \qquad (1)$$

$$\hat{W} = \Re\{\boldsymbol{p} \cdot \boldsymbol{r} e^{is\omega t}\}$$

If $[\hat{W}, \hat{X}] \neq 0$ then $\hat{X}$ is no longer a symmetry of the system. However, regardless of the breaking of $\hat{X}$, the Hamiltonian $\hat{H}$ exhibits the symmetry $\hat{X} \cdot \check{\zeta}_{\hat{X}}$, where $\check{\zeta}_{\hat{X}}$ operates in the synthetic $\boldsymbol{p}$ space. The operation $\check{\zeta}_{\hat{X}}$ may be systematically derived by solving the equation $\hat{W} = \hat{X} \cdot \check{\zeta}_{\hat{X}} \hat{W}$. For the exemplary scenario we focus on here, this translates to

$$\widehat{W} = \Re\{\check{\zeta}_{\hat{X}}(\boldsymbol{p}) \cdot \widehat{X}(re^{is\omega t})\} \tag{2}$$

In Table 1, we employ Floquet group theory[8] and tabulate synthetic-dimension operations $\check{\zeta}_{\hat{X}}$ that solve Eq.(2). The symmetry $\hat{X} \cdot \check{\zeta}_{\hat{X}}$ results in selection rules on various physical observables. For example, in the context of HHG the selection rule for the emitted harmonic light (denoted by $\boldsymbol{E}_{HHG}(\boldsymbol{p},t)$) can be conveniently represented in the time domain by $\boldsymbol{E}_{HHG}(\boldsymbol{p},t) = \hat{X}\boldsymbol{E}_{HHG}(\check{\zeta}(\boldsymbol{p}),t)$. This condition leads to selection rule on the expansion coefficients $E_{nx}^{(klhj)}, E_{ny}^{(klhj)}$ of the n$^{th}$ harmonic amplitude $\boldsymbol{E}_n(\boldsymbol{p})$ (see section I in the SI):

$$\boldsymbol{E}_{HHG}(\mathbf{p},t) = \sum_n \boldsymbol{E}_n(\mathbf{p})e^{in\omega t} \tag{3}$$

$$\boldsymbol{E}_n(\mathbf{p}) \equiv \sum_{k,l,h,j=0}^{\infty} \begin{pmatrix} E_{nx}^{(klhj)} \\ E_{ny}^{(klhj)} \end{pmatrix} p_x^k p_y^l \bar{p}_x^h \bar{p}_y^j$$

Here, $\bar{p}_{x,y}$ are complex conjugates of $p_{x,y}$ respectively, and $k,l,h,j$ are non-negative integers. The symmetry $\hat{X} \cdot \check{\zeta}$ results in selection rules on $E_{nx/y}^{(klhj)}$ to all orders in $k,l,h$ and $j$, and thus provides a non-perturbative formula for the scaling of the HHG spectrum, which are laid out in Table 1. For and expanded discussion on the derivation and application of $\hat{X} \cdot \check{\zeta}$ and its selection rules, see SI, section III. Similar rules were derived for the ATI spectrum (see SI, section II). We further emphasize that the only necessary condition for this construction is that the system exhibits a broken-symmetry, and that Eq. (1) holds. For example, in section V of the SI we derive $\hat{X} \cdot \check{\zeta}$ for a system whose real-symmetry $\hat{X}$ is broken by spin-orbit coupling, instead of the second laser (i.e., $\widehat{W}$ is different). In this case, $\check{\zeta}$ operates on the synthetic space $(\alpha, \gamma)$ spanned by the Rashba $(\alpha)$ and Dresselhaus $(\gamma)$ coupling strengths.

Next, we demonstrate experimentally synthetic symmetries and their corresponding selection rules in a symmetry-broken system by driving HHG with a bi-chromatic field composed of incommensurate frequencies. In our set-up[27] (Fig.1.(a)), a bi-chromatic laser beam with frequencies $\omega - 1.95\omega$ (corresponding to the wavelengths 800nm and 410nm, respectively) is passed through an achromatic zero-order quarter-wave plate (QWP). The rotation angle θ of the QWP controls the ellipticities (and the DSs) of the driving field, which is explicitly given by

$$\mathbf{E}(t) = \left(\cos(\omega t)\hat{\mathbf{x}} + \sin(1.95\omega t)\hat{\mathbf{y}}/\sqrt{10}\right) \quad (4)$$
$$+ \sin(2\theta)\left(-\cos(1.95\omega t)\hat{\mathbf{x}}/\sqrt{10} + \sin(\omega t)\hat{\mathbf{y}}\right)$$
$$+ \cos(2\theta)\left(\sin(\omega t)\hat{\mathbf{x}} + \cos(1.95\omega t)\hat{\mathbf{y}}/\sqrt{10}\right)$$

The bi-chromatic beam is focused onto a supersonic jet of argon gas, yielding an intensity of $2 \times 10^{14}$ W/cm² at the focus, where 10% of the intensity is in the redshifted SH driver. The measured HHG spectrum (Fig.1. (b)) exhibits two types of selection rules, imposed by standard and synthetic DSs.

Firstly, for $\theta = 0°$ and $\theta = 45°$, the driving field exhibits standard HHG selection rules in the form of forbidden harmonics, due to the spatio-temporal DSs $\hat{Z}'_x = \hat{\tau}'_2 \cdot \hat{\sigma}_x$ and $\hat{C}'_{59,20} = \hat{\tau}'_{20} \cdot \hat{R}_{59,20}$, respectively. Here, $\hat{\tau}'_N$ is a $T'/N$ time translation where $T' = 20T$, $\hat{\sigma}_x$ is a reflection relative to the Cartesian basis vector $\hat{x}$, and $\hat{R}_{59,20}$ is a $20 \times 2\pi/59$ spatial rotation. Hence[8], even harmonic generation is forbidden for $\theta = 0°$ and $3n$ harmonic generation is forbidden for $\theta = $ 45° (Fig.1.(b)). As $\theta$ is detuned from 0° and 45°, the DSs $\hat{Z}'_x$ and $\hat{C}'_{59,20}$ are broken, and instead, DSs of the form $\hat{Z}'_x \cdot \check{\zeta}$ and $\hat{C}'_{59,20} \cdot \check{\zeta}$ are imposed, where $\check{\zeta}$ are synthetic dimension operations given by Table 1. In the following, we demonstrate that the $\theta$-scaling of the spectrum is controlled by the selection rules corresponding to these synthetic symmetries, for three exemplary spectral components of the emission – H18.8, H19.75 and H20.7 (where H stands for harmonic of the fundamental $\omega$ pump). As θ is detuned from 0°, the $\hat{Z}'_x$ DS is broken (to 1st order in θ) by two linearly polarized fields with frequencies $\omega \equiv 20\omega'$ and $1.95\omega = 39\omega'$, given explicitly by $(-2\theta \cos(39\omega' t) \hat{x}/\sqrt{10})$ and $(+2\theta \sin(20\omega' t) \hat{y})$ (these 2nd perturbative fields can be thought of as arising due to the changing ellipticity of the original beams when the QWP is rotated – any deviation from the original Floquet Hamiltonian can be assigned to a perturbative term). By plugging in either $(p_x = 2\theta, p_y = 0, s = 39)$ or $(p_x = 0, p_y = 2\theta, s = 40)$ into Table 1 and employing $\omega'$ as the fundamental frequency, we identify that the system exhibits $\hat{Z}'_x \cdot \check{\zeta}$ symmetry where $\check{\zeta}(\theta) = -\theta$ (Fig.1.(c)). The corresponding selection rules forbids the amplitudes of harmonic H18.8 and H20.7 to have a contribution proportional to $p_x = 2\theta$ in their scaling, hence, they scale quadratically, as $\theta^2$. Similarly, $\hat{Z}'_x \cdot \check{\zeta}$ forbids H19.75 to have a contribution proportional to $p_x^2 \propto \theta^2$, hence it scales linearly with $\theta$. See section VI of the SI for an explicit derivation of these predictions using Table 1. To compare the scaling of each harmonic amplitude to the analytical predictions, we fitted it to a general polynomial model (Fig.1(e-g), yellow shaded regions). The ratio between the coefficients of the polynomial fit indicates whether the emission scales linearly or quadratically, e.g., in Fig.1.(g), H19.75 scales as $9.5(\theta - 0.1) + 0.05(\theta - 0.1)^2$ in the yellow-shaded region, in accordance with the analytically derived selection rule that forbids quadratic ($\propto \theta^2$) contributions (because the coefficient for $\theta^2$ is ~200 times weaker than the $\theta$ coefficient). Additionally, it was verified that when fitting the scaling of each harmonic amplitude to its specific prediction, i.e., to a linear or quadratic power-law, all fits result in $R^2>0.95$.

As θ is detuned away from 45°, we probe the breaking of $\hat{C}'_{59,20}$ DS (to 1st order in θ) by two linearly polarized fields $\sin(20\omega't)\,\hat{x}$ and $\cos(39\omega't)/\sqrt{10}\,\hat{y}$ with amplitude $2(\theta°-45°)$. As a result, the $\hat{C}'_{59,20}$ is broken, and synthetic symmetries of the form $\hat{C}'_{59,20}\cdot\check{\zeta}$ are imposed (Fig.1. (d)). To obtain the corresponding selection rules in terms of the detuning angle $(\theta-45°)$, we plug in $(p_x \propto (\theta-45°), s=20)$ and $(p_y \propto (\theta-45°), s=39)$ into Table 1 (see SI section VI for an explicit substitution). The corresponding selection rules forbids contributions linear in $\theta$ to the scaling of H20.7 and H19.75. Additionally, it forbids linear, quadratic, and cubic contributions, to the scaling of H18.8, hence it is predicted to scale as a quartic function of the detuning angle $(\propto (\theta-45°)^4)$. These predictions (Fig.1. (e)) are clearly observed in the experimental measurement (Fig.3(f-h), blue shaded regions).

To summarize, we have demonstrated that systems that are traditionally regarded as symmetry-broken, can systematically exhibit a new class of symmetries and selection rules through synthetic dimensions. These symmetries and selection rules determine how the system's observables scale as the system transitions out of its original symmetric state, showing the role of the broken symmetries in the dynamics of the symmetry-broken systems. We have tabulated these symmetries for periodically driven Floquet systems subject to two driving fields, and derived the corresponding selection rules for HHG, ATI and more (see section VII in the SI). We observed experimentally selection rules rooted in synthetic dimensions by driving HHG with a bi-chromatic laser field that consists of incommensurate frequencies. We highlight that our theory is a non-perturbative theory, that applies to all orders of the perturbation's strength. We further emphasize that real-synthetic symmetries and their associated selection rules are general concepts, relevant to all systems with a broken symmetry in real space. For example, one (or both) of the lasers we have employed in our construction, may be replaced by a different periodically oscillating (or static) element of the system (either extrinsic or intrinsic), e.g., spin orbit coupling strengths[28] (see section V in the SI) or lattice excitations[25,26,29]. Specifically, by reformulating them as effective gauge fields, the derived symmetries and selection rules (Table 1) can be directly applied to dynamical symmetry breaking by phonons and magnons, opening new opportunities for all-optical time-resolved spectroscopy (and control) of their dynamics. Overall, the presented approach provides a unified framework for the analysis of symmetry-broken systems, complementary to perturbation theory, hence we expect it to be used throughout science and engineering.

**Table I: Synthetic dynamical symmetries and their corresponding selection rules for harmonic generation and photoionization.** When the DS is not $\hat{C}_{NM}$ or $\hat{e}_{NM}$, the coefficients $\left(k, h, j, l, E_{nx}^{(klhj)}, E_{ny}^{(klhj)}\right)$ are defined in the Eq. (3). For $\hat{C}_{NM}$ & $\hat{e}_{NM}$, the expansion coefficients $(E_{Rn}^{(klhj)}, E_{Ln}^{(klhj)}, k, l, h, j)$ are defined in the SI, section I. The coefficients $\tilde{\phi}_{kn}^{(abcd)}$ and their selection rules were defined & derived in the SI, section II.

| $\hat{X}$ | $\check{\zeta}_s(\boldsymbol{p})$ | Harmonic generation selection rule | Above threshold ionization selection rule |
|---|---|---|---|
| $\hat{T}$ | $\bar{\boldsymbol{p}}$ | $E_{nx}^{(klhj)}, E_{ny}^{(klhj)} \in \mathbb{R}$ | $\tilde{\phi}_{kn}^{(abcd)} = \tilde{\phi}_{k,-n}^{(cdab)}$ |
| $\hat{Q}$ | $-\bar{\boldsymbol{p}}$ | $E_{nx}^{(klhj)}, E_{ny}^{(klhj)} \in i^{1+k+l+h+j}\mathbb{R}$ | $\tilde{\phi}_{kn}^{(abcd)} = (-1)^{a+b+c+d}\tilde{\phi}_{(-k)(-n)}^{(cdab)}$ |
| $\hat{G}$ | $(-1)^{1+s}\bar{\boldsymbol{p}}$ | $E_{nx}^{(klhj)}, E_{ny}^{(klhj)} \in i^{n+1+(s+1)(k+l+h+j)}\mathbb{R}$ | $\tilde{\phi}_{kn}^{(abcd)} = \tilde{\phi}_{(-k)(-n)}^{(cdab)}(-1)^{n+(s+1)(a+b+c+d)}$ |
| $\hat{Z}_y$ | $\begin{pmatrix}(-1)^{s+1} & 0 \\ 0 & (-1)^s\end{pmatrix}\begin{pmatrix}p_x \\ p_y\end{pmatrix}$ | $n + (s+1)(k+h) + s(j+l) = 2q \Rightarrow E_{nx}^{(klhj)} = 0$ | $\tilde{\phi}_{\alpha,k_n}^{(abcd)} = \tilde{\phi}_{\alpha,\hat{\sigma}_y k_n}^{(abcd)}(-1)^{n+(s+1)(a+c)+s(b+d)}$ |
| | | $n + (s+1)(k+h) + s(j+l) = 2q+1 \Rightarrow E_{ny}^{(klhj)} = 0$ | |
| $\hat{D}_y$ | $\begin{pmatrix}-1 & 0 \\ 0 & 1\end{pmatrix}\begin{pmatrix}\bar{p}_x \\ \bar{p}_y\end{pmatrix}$ | $E_{nx}^{(klhj)} \in i^{1+k+h}\mathbb{R}$ $E_{ny}^{(klhj)} \in i^{k+h}\mathbb{R}$ | $\tilde{\phi}_{kn}^{(cdab)} = \tilde{\phi}_{(\hat{\sigma}_y k)(-n)}^{(abcd)}(-1)^{(a+c)}$ |
| $\hat{H}_y$ | $\begin{pmatrix}(-1)^{s+1} & 0 \\ 0 & (-1)^s\end{pmatrix}\begin{pmatrix}\bar{p}_x \\ \bar{p}_y\end{pmatrix}$ | $E_{nx}^{(klhj)} \in i^{n+1+(s+1)(k+h)+s(l+j)}\mathbb{R}$ $E_{ny}^{(klhj)} \in i^{n+(s+1)(k+h)+s(l+j)}\mathbb{R}$ | $\tilde{\phi}_{kn}^{(abcd)} = \tilde{\phi}_{\sigma_y k,-n}^{(cdab)}(-1)^{n+(s+1)(a+c)+s(b+d)}$ |
| $\hat{C}_{NM}$ | $e^{-\frac{i2\pi s}{N}}\hat{R}^{(p)}_{N,M}\cdot\boldsymbol{p}$ | $E_{Rn}^{(klhj)}(p) \neq 0$ if $\exists q \in \mathbb{N}$: $n - M(l+h-k-j) - s(k+l-h-j) + M = Nq$ | $\tilde{\phi}_{k_n}^{(abcd)} = \tilde{\phi}_{\hat{R}_{NM}k_n}^{(abcd)} e^{i\frac{2\pi}{N}[n+(M-s)(a-c)-(M+s)(b-d)]}$ |
| | | $E_{Ln}^{(klhj)}(p) \neq 0$ if $\exists q \in \mathbb{N}$: $n - M(l+h-k-j) - s(k+l-h-j) - M = Nq$ | |
| $\hat{e}_{NM}$ | $e^{-\frac{i2\pi s}{N}}\hat{L}^{(p)}_{1/b}\cdot\hat{R}^{(p)}_{N,M}\cdot\hat{L}^{(p)}_b\cdot\boldsymbol{p}$ | $E_{-n}^{(klhj)}(p) \neq 0$ if $\exists q \in \mathbb{N}$: $n - M(l+h-k-j) - s(k+l-h-j) + M = Nq$ | $\tilde{\phi}_{k_n}^{(abcd)} = \tilde{\phi}_{\hat{L}_b\cdot\hat{R}_{N,M}\cdot\hat{L}_{1/b}k_n}^{(abcd)} e^{i\frac{2\pi}{N}[n+(M-s)(a-c)-(M+s)(b-d)]}$ |
| | | $E_{+n}^{(klhj)}(p) \neq 0$ if $\exists q \in \mathbb{N}$: $n - M(l+h-k-j) - s(k+l-h-j) - M = Nq$ | |

**Figure 1: Experimental investigation of selection rules by real-synthetic symmetries.** (a) illustration of the experimental setup (b) measured HHG spectrum as a function of the QWP angle $\theta$. (c) Illustration of the symmetry operations $\hat{Z} \cdot \check{\zeta}_{\hat{Z}}$ on the Lissajous curves of the driving field at $\theta = 0°$ and $\theta = 10°$. (d) Illustration of the symmetry operation $\hat{C} \cdot \check{\zeta}_{\hat{C}}$ on the Lissajous curve of the driving field for $\theta = 45°$ and $\theta = 55°$. The axes $p_x$ and $p_y$ in the x and y amplitude components of the symmetry breaking part of the field, i.e., they are proportional to $(\theta - 45°)$. (e) summary of analytical predictions for the scaling of each spectral component (f-h) The scaling of harmonics 18.8, 19.75 and 20.7. The scaling of each harmonic amplitude was fit to a general polynomial model within the yellow & blue shaded regions, and the resulting polynomials appear in the corresponding color above each subfigure. The ratio between the coefficients of the polynomial indicates whether the emission scales linearly or quadratically.

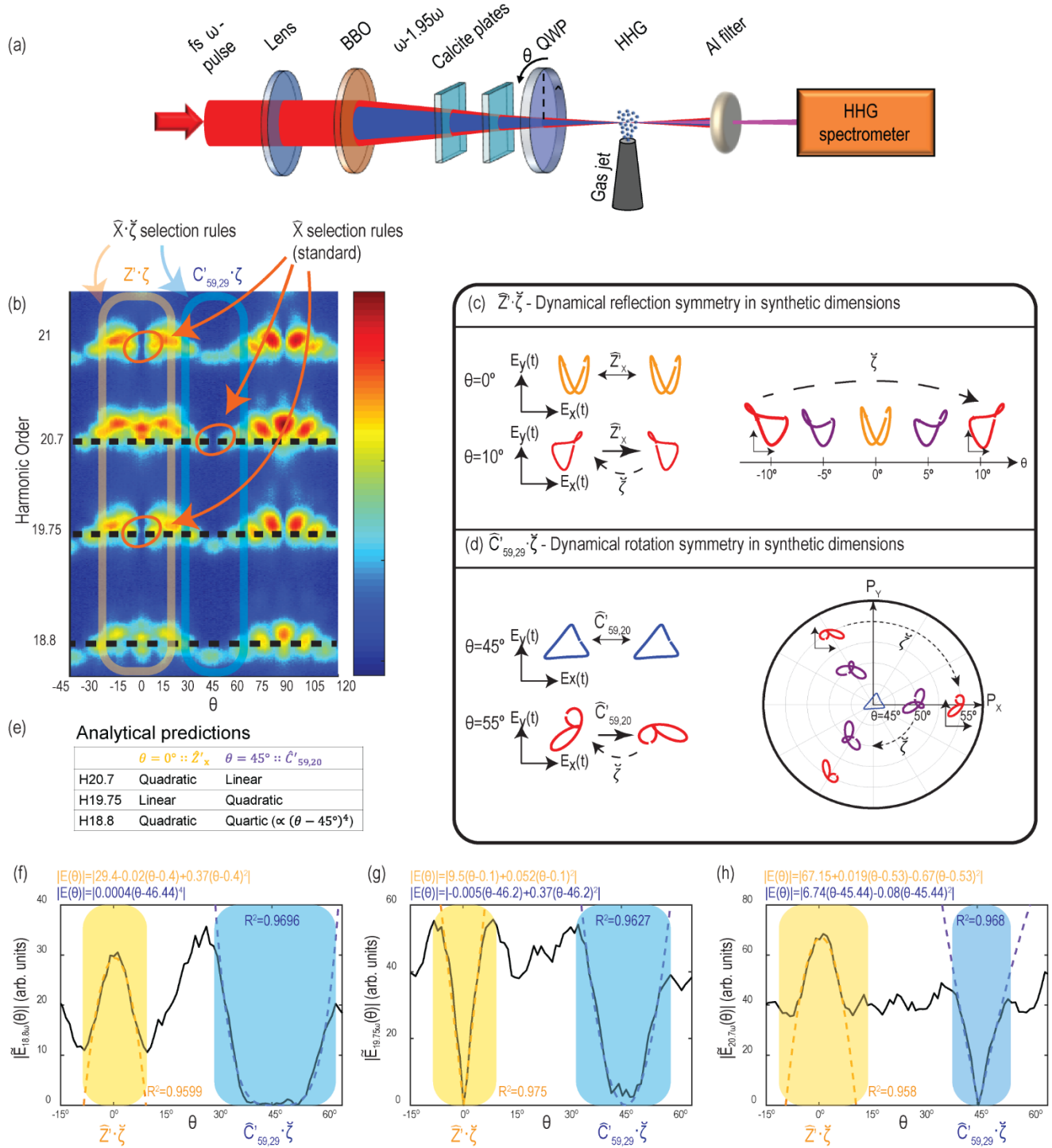

## Data Availability Statements

The data supporting the findings of this study are available from the corresponding author upon reasonable request.

## Acknowledgements

## Contributions

All authors made substantial contributions to all aspects of the work.

## Competing interests

The authors declare no competing interest